\documentclass{elsart}
\usepackage{epsfig}
\usepackage{graphicx, natbib, amssymb}
\journal{New Astronomy}
\begin{document}
\begin{frontmatter}
\title{A family of triaxial
 modified Hubble mass models: effects of the 
additional radial functions}
\author[Korea]{Mousumi Das},
\ead{mdas@pusan.ac.kr}
\author[Korea]{Parijat Thakur\corauthref{cor}},
\corauth[cor]{Corresponding author.}
\ead{pthakur@pusan.ac.kr}
\author[Korea]{H.B. Ann},
\ead{hbann@pusan.ac.kr}
\address[Korea]{Division of Science Education, Pusan National University,
 Busan 609-735, Korea}

\begin{abstract}

The projected properties of triaxial generalization of 
the modified Hubble mass models are studied. These models 
are constructed by adding the additional radial functions, 
each multiplied by a low-order spherical harmonic, to 
the models of \citet{ct00}. The projected surface density 
of mass models can be calculated analytically which allows us 
to derive the analytic expressions of axial ratio and position 
angles of major axis of constant density elliptical contours 
at asymptotic radii. The models are more general than 
those studied earlier in the sense that the inclusions of 
additional terms in density distribution, allows one to 
produce varieties of the radial profile of axial ratio and 
position angle, in particular, their 
small scale variations at inner radii. Strong correlations are found to 
exist between the observed axial ratio evaluated at $0.25 R_{e}$
and at $4 R_{e}$ which occupy well-separated regions in 
the parameter space for  different choices of the intrinsic 
axial ratios. These correlations can be exploited to predict
the intrinsic shape of the mass model, independent of 
the viewing angles. Using Bayesian 
statistics, the result of a test case launched for an 
estimation of the shape of a model galaxy is found to be satisfactory. 
\end{abstract}
\begin{keyword} 
galaxies : triaxial - galaxies : photometry - galaxies : structure.
\end{keyword}
\end{frontmatter}
 
\section{Introduction}
Observed photometric properties of majority of elliptical galaxies
 show isophotal twists and variations in ellipticity and position 
angle of major axis with radius. These observed properties can be 
produced in a natural way by triaxial models.
\citet[][hereafter CT00]{ct00} studied the projected
properties of a family of mass models which are  triaxial
 generalisation of modified Hubble mass model. The model 
was first proposed by \citet{Sch79} as a numerical model for 
a triaxial stellar system in dynamical equilibrium and 
later casted into analytical form by 
\citet{DeM}. It was found by CT00 that these models show 
ellipticity variations and isophotal twists in
their projections along the line of sight. Moreover,
the radial profiles of the parameters of the elliptical isophotes 
were found to be smooth functions of the radius (see CT00). 
However, many elliptical galaxies, devoid of any features, 
indicating the absence of shells or dust, are found to 
exhibit small scale variations in the radial profiles of 
the parameters of elliptical isophotes. This indicates 
that density distributions of such ellipticals may be more 
complex than that of CT00. 

Above fact inspired us to modify the models of CT00 by adding
 the additional radial functions, each multiplied by a low-order 
spherical harmonic. The projected surface density $\Sigma$ of 
resultant mass model can be calculated analytically. This makes 
 possible to investigate some of the projected properties 
analytically. We calculated the profiles of surface density
$\Sigma$, axial ratio $b/a$ and position angle $\Theta_{*}$ 
of the major axis as functions of radius. The analytical
 expression of the $\Sigma$ is very useful for 
the calculations of $b/a$ and $\Theta_{*}$. In the asymptotic 
regions analytical expressions can be derived, while in 
the intermediate regions simple numerical methods can be 
adopted for these calculations. The inclusions of additional terms 
in the density distribution, allows us to produce 
varieties of the radial profile of  $b/a$ and $\Theta_{*}$ that
can be compared with photometric data of real galaxies.

It was shown by \citet{DeC} that the observed ellipticities and
the position angle, at small and at large radii can be used to 
derive the intrinsic axial ratios of the mass model, as a function 
of the viewing angles. Since the viewing angles of galaxies are 
largely unknown, it will be worth-while to get an estimate of 
the intrinsic axial ratios independent of the viewing angles. 
In this regard, \citet[][hereafter SF94]{stat94}, 
and \citet[][hereafter TC01]{tdkc} found that in case the
 observed parameters exhibit correlations when a model with a
 given set of intrinsic parameters is viewed in all possible
 orientations, one can obtain a probable estimate of the 
intrinsic shape, independent of the viewing angles.
Furthermore, it was also shown by TC01 that the intrinsic
shapes of triaxial mass models can be estimated using photometric 
properties and further, the results of shape estimation 
would be insensitive to the choice of models, if one considers 
ensembles of models which represent above mentioned correlations 
between the observed parameters. Thus, another aim of our present
study is to build ensembles of models showing the correlations 
between observed parameters so that one could get an estimate
of model independent shape.

In \S 2, we describe the mass model and in \S 3, we
 present the projected properties. Our results are described 
in \S 4, and \S 5 is devoted to summary and
 discussion.

\section{The mass model }
We modified the triaxial potential, given in \citet{DeM}, corresponding 
to the density
distribution considered in CT00 by adding 
the additional radial functions $v_{e}(r)$ and $w_{e}(r)$, 
each multiplied by a low-order spherical harmonic.
Thus, our assumed potential $V(r, \theta, \phi)$ has a following form
\begin{eqnarray}
V(r, \theta, \phi) & = & u(r)-(v(r)+v_{e}(r)) Y_{2}^{0}(\theta)+
(w(r)+w_{e}(r)) Y_{2}^{2}(\theta,\phi), \
\end{eqnarray} 

where $u(r)$, $v(r)$, $w(r)$, $v_{e}(r)$ and $w_{e}(r)$ are five 
radial functions, $(r,\theta,\phi)$  are spherical coordinates
 defined such that $x=r\sin\theta\cos\phi$,  $y=r\sin\theta\sin\phi$ 
and $z=r\cos\theta$, and $Y_{2}^{0}(\theta)  = 
\frac {3} {2} cos^{2}\theta-\frac {1} {2}$ and $Y_{2}^{2}(\theta,\phi)
 = 3 sin^{2}\theta cos2\phi$ are the usual spherical harmonics.

We take $u(r)$ to be the 
potential of the spherical modified Hubble mass model, defined by 
\begin{eqnarray}
u(r) &=& - GM\frac{ln[r+\sqrt{b_{o}^{2}+r^{2}}]}{r} \ ,
\end{eqnarray} 
where $M$ is the total mass of the model and $b_{o}$ is 
the scale length.
 We choose $v(r)$ and $w(r)$ to be same as adopted in \citet{DeM} and 
reproduced below with slightly different notations.

\begin{eqnarray}
v(r) &=& -\frac{GM}{b_{o}^{3}}\ \frac{b_{1}^{3}r^{2}}{(b_{2}^{2}+
r^{2})^{3/2}} \nonumber\\
w(r) &=& -\frac{GM}{b_{o}^{3}}\ \frac{b_{3}^{3}r^{2}}{(b_{4}^{2}+
r^{2})^{3/2}},
\end{eqnarray} 
where $b_{1}, ..., b_{4}$ are constants. The additional radial
functions $v_{e}(r)$ and $w_{e}(r)$ are chosen such that they 
are effective in intermediate range of $r$ only and not 
at small and at large radii. In order to satisfy this 
condition, a suitable choice of $v_{e}(r)$ and $w_{e}(r)$ are   
\begin{eqnarray}
v_{e} (r) &=& -\frac{GM}{b_{o}^{3}} \ \frac{a_{1}^{4}r^{6}}{(a_{2}^{2}+
r^{2})^{4}}  \nonumber\\
w_{e} (r) &=& -\frac{GM}{b_{o}^{3}} \ \frac{a_{3}^{4}r^{6}}{(a_{4}^{2}+
r^{2})^{4}},
\end{eqnarray}
where $a_{1}, ..., a_{4}$ are constants. We note that 
$v(r)$ and $w(r)$ go as $-r^{2}$ at small radii and as 
$-1/r$, at large radii. On the other hand, $v_{e}(r)$ and $w_{e}(r)$ 
go as $-r^{6}$ at small radii and as $-1/r^{2}$ at large radii.

For the above considered potential, the associated density 
distribution $\rho(r,\theta,\phi)$ follows from Poisson's 
equation  
\begin{eqnarray}
\rho(r, \theta, \phi) & = & f(r)-(g(r)+g_{e}(r))\  
Y_{2}^{0}(\theta)+ (h(r)+h_{e}(r)) \  Y_{2}^{2}(\theta,\phi).
\label{eq:rhoe} 
\end{eqnarray}
The five radial functions appeared in equation (\ref{eq:rhoe}) are
 given by
\begin{eqnarray}
f(r) & = & \frac {M} {4 \pi } \ \frac {1} 
{(b_{o}^{2}+r^{2})^{3/2}}, \nonumber \\
g(r) & = &\frac {3M} {4 \pi } \ \frac {b_{1}^{3}} {b_{o}^{3}} \ 
\frac {2 r^{4}+7 b_{2}^{2} r^{2}} {(b_{2}^{2}+r^{2})^{7/2}}, 
\nonumber \\
g_{e}(r) & = &\frac {M} {4 \pi} \ \frac{4a_{1}^{4}}{b_{o}^{3}}\ 
\frac {r^{8}+12a_{2}^{2}r^{6}-9a_{2}^{4}r^{4}}
 {(a_{2}^{2}+r^{2})^{6}}, \nonumber \\
h(r) & = &\frac {3M} {4 \pi }\frac { b_{3}^{3}} {b_{o}^{3}}\ 
\frac {2 r^{4}+7 b_{4}^{2} r^{2}} {(b_{4}^{2}+r^{2})^{7/2}}, 
\nonumber \\
h_{e}(r) & = &\frac {M} {4 \pi } \ \frac{4a_{3}^{4}}{b_{o}^{3}} 
\frac {r^{8}+12a_{4}^{2}r^{6}-9a_{4}^{4}r^{4}} 
{(a_{4}^{2}+r^{2})^{6}}. 
\end{eqnarray}
The radial dependence of additional 
functions $g_{e}(r)$ and $h_{e}(r)$ in the density distribution are 
such that at asymptotic radii, the density remains same as 
that of CT00. At large radii, $g_{e}(r)$ and $h_{e}(r)$ 
decrease as $r^{-4}$, whereas $g(r)$ and $h(r)$ decrease 
as $r^{-3}$. Likewise, at small radii, $g_{e}(r)$ and $h_{e}(r)$ 
decrease as $r^{4}$, whereas $g(r)$ and $h(r)$ goes
as $r^{2}$. Thus, the radial functions $f(r)$, $g(r)$ and 
$h(r)$ are dominating at small and at large radii in 
the density distribution (\ref{eq:rhoe}). The four ratios 
$(b_{1}/b_{o})$, ... , $(b_{4}/b_{o})$, appeared in $g(r)$ and 
$h(r)$, are expressed in terms of axial ratios of the
approximately ellipsoidal constant $\rho$ surfaces at small and 
at large radii by the same way as adopted in CT00. 
Prescribing the values of axial ratios of the
approximately ellipsoidal constant $\rho$ 
surfaces at very large and at very small radii as 
$(p_{\infty}, q_{\infty})$ and  $(p_{o}, q_{o})$, respectively, 
we find that 
\begin{eqnarray}
\left(\frac{b_{1}}{b_{0}}\right)^{3} &=& 
\frac{\left(1+p_{\infty}^{3}-2q_{\infty}^{3}\right)}
{6\left(1+p_{\infty}^{3}+q_{\infty}^{3}\right)} \ , \nonumber \\
\left(\frac{b_{3}}{b_{0}}\right)^{3} &=& 
\frac{\left(1-p_{\infty}^{3}\right)}{12\left(1+p_{\infty}^{3}
+q_{\infty}^{3}\right)} \ , \nonumber \\
\left(\frac{b_{2}}{b_{0}}\right)^{5} &=& 
\left[\frac{1+p_{\infty}^{3}-2q_{\infty}^{3}}
{6\left(1+p_{\infty}^{3}+q_{\infty}^{3}\right)}\right]
\left[\frac{7\left(p_{0}^{2}+4q_{0}^{2}-1\right)}
{1+p_{0}^{2}-2q_{0}^{2}}\right],\nonumber\\
\left(\frac{b_{4}}{b_{0}}\right)^{5} &=& 
\left[\frac{1-p_{\infty}^{3}}{12\left(1+p_{\infty}^{3}
+q_{\infty}^{3}\right)}\right]
\left[\frac{42\left(p_{0}^{2}+4q_{0}^{2}-1\right)}{2-6q_{0}^{2}}\right].
\end{eqnarray}
Here it is clear that the expressions of $(b_{1}/b_{o})$ and 
$(b_{3}/b_{o})$ are exactly similar to those of CT00, whereas 
they are found to be completely different for $(b_{2}/b_{o})$ 
and $(b_{4}/b_{o})$. Moreover, throughout the paper, 
the constant parameters $a_{1}, ..., a_{4}$ are chosen such that 
$|g_{e}(r)| \ll |g(r)|$ and  $|h_{e}(r)| \ll |h(r)|$ at all $r$. 
\section{Projected properties }
\subsection{Projected surface density}
The density form (\ref{eq:rhoe}) of the models allows a straight 
forward calculation of the projected surface density $\Sigma$ for any
 viewing angle. We adopted the convention of \citet{DeF} to project
the model along line-of-sight. We chose coordinates 
$(x^{'},y^{'},z^{'})$ such that $z^{'}$-axis runs along the 
line-of-sight, and $x^{'}$-axis lies in the $(x,y)$ plane. We 
considered $(\theta^{'},\phi^{'})$ as the standard spherical
coordinates of the line-of-sight and  $(R,\Theta)$ as the 
polar coordinates in $(x^{'},y^{'})$ plane. The projected 
surface density $\Sigma(R,\Theta)$ is obtained by integrating 
the model density along the line-of-sight,
\begin{eqnarray}
\Sigma(R,\Theta) &=& \int_{-\infty}^{+\infty} \rho dz^{'},
\end{eqnarray}
and after doing some intermediate calculations, it takes the 
following form
\begin{eqnarray}
\Sigma\left(R,\Theta\right) & = & \Sigma_{o}(R)  +
 \Sigma_{2}(R) \ \cos \ 2\left(\Theta - \Theta_{*}\right),
\label{eq:sige}
\end{eqnarray}
where 
\begin{eqnarray}
\Sigma_{o} \left(R\right) & = & 2F_{1} + 
\left(1-3 \cos^{2}\theta^{'}\right) \ 
\left[(G_1+G_{1e}) - \frac{3} {2}(G_2+G_{2e})\right] +{}\nonumber\\
& & + {} \left[6(H_1 +H_{1e})- 9(H_2+H_{2e})\right]
 \ \sin^2\theta^{'} \cos 2\phi^{'},\nonumber\\
\Sigma_{2}^{2}(R) & = & \left[6 \ (H_2+H_{2e}) \ 
\cos\theta^{'} \ \sin2\phi^{'}\right]^{2} +
 \left[\frac{3} {2}(G_{2}+G_{2e})\sin^2\theta^{'} -{} \right.\nonumber \\
& & -{}\left. 3(H_{2}+H_{2e})\left(1 + 
\cos^2\theta^{'}\right)\ \cos2\phi^{'} \right]^{2}.
\end{eqnarray}

We have defined the integrals
\begin{displaymath}
G_1(R) = \int^{\infty}_{R} \frac{r \ g(r) \ dr} 
{\sqrt{\left(r^2 -R^2\right)}} \ ,\qquad G_{1e}(R) = \int^{\infty}_{R} 
\frac{r \ g_{e}(r) \ dr} {\sqrt{\left(r^2 -R^2\right)}}, \nonumber 
\end{displaymath}

\begin{eqnarray}
G_2(R) = R^{2} \ \int^{\infty}_{R} \frac { \ g(r) \ dr} {r \ 
\sqrt{\left(r^2 -R^2\right)}} \ ,\qquad G_{2e}(R) = R^{2} \ 
\int^{\infty}_{R} \frac { \ g_{e}(r) \ dr} {r \ 
\sqrt{\left(r^2 -R^2\right)}},
\end{eqnarray}
and similarly integrals for $F_{1}$, $H_{1}$, $H_{2}$, $H_{1e}$
and $H_{2e}$ in terms of functions $f(r)$, $h(r)$ and $h_{e}(r)$, 
respectively.

Above integrals are calculated analytically and can be expressed as
\begin{eqnarray}
F_{1}(R) & = & \frac {M} {4 \pi b_{o}^{3}} \ 
\frac {b_{o}^{3}} {b_{o}^{2}+R^{2}} \nonumber , \\
G_{1}(R) & = & \frac {M} {4 \pi b_{o}^{3}} \
 \frac {2 b_{1}^{3}} {\left(b_{2}^{2}+R^{2}\right)^{3}} \
 \left[2b_{2}^{4}+9b_{2}^{2} R^{2}+3R^{4}\right] \nonumber ,  \\
G_{1e}(R) & = & \frac {M} {4 \pi b_{o}^{3}}\ 
\left[4a_{1}^{4}I_{3}+12 a_{2}^{2}I_{2}-9a_{2}^{4}I_{1}\right]\nonumber , \\
G_{2}(R) & = & \frac{M} {4 \pi b_{o}^{3}} \ 
\frac {4b_{1}^{3}} {\left(b_{2}^{2}+R^{2}\right)^{3}} R^{2} \ 
\left[3b_{2}^{2}+R^{2}\right] \ \nonumber , \\
G_{2e}(R) & = & \frac {M} {4 \pi b_{o}^{3}}\ 
\left[4a_{1}^{4}R^{2}I_{2}+12 a_{2}^{2}I_{1}-9a_{2}^{4}I_{4}\right],
\label{eq:fghe}
\end{eqnarray}
where 
\begin{displaymath}
a_{2}^{2} + R^{2} =\alpha^{2} \ \ \
 \ \ \ \  \beta = \frac{1}{512 \ \alpha^{11}},
\end{displaymath}
\begin{eqnarray}
I_{1} &=& \beta (63R^{4} + 14R^{2}\alpha^{2} +3\alpha^{4})\nonumber , \\
I_{2} &=& \beta (63R^{6} + 21R^{4}\alpha^{2} +9R^{2}\alpha^{4} 
+3\alpha^{6})\nonumber , \\
I_{3} &=& \beta (63R^{8} + 28R^{6}\alpha^{2} +18R^{4}\alpha^{4} 
+12R^{2}\alpha^{6}+7\alpha^{8})\nonumber , \\
I_{4} &=& \beta (63R^{2}+7\alpha^{2}). 
\end{eqnarray}
Similarly, $H_{1}(R)$ and $H_{2}(R)$ can be written by 
substituting $b_{3}$ and $b_{4}$ in place of $b_{1}$ 
and $b_{2}$ in $G_{1}(R)$ and $G_{2}(R)$, respectively.
On the other hand, $H_{1e}(R)$ and $H_{2e}(R)$ are defined by 
substituting $a_{3}$ and $a_{4}$ in place of $a_{1}$ 
and $a_{2}$ in $G_{1e}(R)$ and $G_{2e}(R)$, respectively.
 
The projected surface density (\ref{eq:sige}) has its 
major axis at position angle $\Theta_{*}$, which is given by
\begin{eqnarray}
\tan2\Theta_{*} & = & \frac { T \ h_{3} } {h_{1} + 
\left(1-T\right) \ h_{2}},
\label{eq:pos}
\end{eqnarray}
where $h_{1}$, $h_{2}$ and $h_{3}$ depend only on the viewing 
angles $(\theta^{'},\phi^{'})$ and are defined as  
\begin{eqnarray}
h_{1} & = & \sin^{2} \phi^{'}-\cos^{2}\phi^{'} \ \cos^{2} 
\theta^{'} \nonumber , \\
h_{2} & = & \cos^{2} \phi^{'}-\sin^{2}\phi^{'} \ \cos^{2} 
\theta^{'} \nonumber , \\
h_{3} & = & \sin \ 2\phi^{'} \ \cos \theta^{'}.
\label{eq:hfun}
\end{eqnarray}
The quantity $T \equiv T(R)$ in equation (\ref{eq:pos}) is 
the triaxiality parameter, which is given by
\begin{eqnarray}
T \equiv T(R) & = & \frac { 4 \ (H_{2}(R)+H_{2e}(R))}
 {(G_2(R)+G_{2e}(R)) + 2 \ (H_{2}(R)+H_{2e}(R))}.
\label{eq:trip} 
\end{eqnarray}
Equation (\ref{eq:pos}) gives the position angle $\Theta_{*}$ of the
major axis when it satisfies the condition
\begin{eqnarray}
(H_{2}+H_{2e})h_{3}\sin 2\Theta_{*}< 0.
\end{eqnarray}
The observed axial ratio $b/a$ of the projected surface density can
 be calculated by using the relation
\begin{eqnarray}
\Sigma(b,\Theta_{*}-\frac{\pi}{2}) &=& \Sigma(a,\Theta_{*}), 
\end{eqnarray}
which can be rewritten as
\begin{eqnarray}
\Sigma_{o}(a)+\Sigma_{2}(a) & = & \Sigma_{o}(b) - \Sigma_{2}(b).
\label{eq:ba}
\end{eqnarray}
At a finite radial distance from the centre, the axial
 ratio can be calculated numerically 
using equation (\ref{eq:ba}). However, at asymptotic limits,
the analytical expressions can be obtained for $b/a$.

\subsection{Projected properties at large and at small radii}

The ratio $(G_{2}+G_{2e})/(H_{2}+H_{2e})$ in 
(\ref{eq:trip}) can be calculated analytically using 
(\ref{eq:fghe}), which in turn enables us to write 
an analytical expressions for $T$ and $\Theta_{*}$. 
In particular, at large $R$, the ratio 
$[(G_{2}+G_{2e})/(H_{2}+H_{2e})]_{\infty}$ and 
the triaxiality parameter $T_{\infty}$ are given by 
\begin{eqnarray}
 \left(\frac {G_2+G_{2e}} {H_2+H_{2e}}\right)_{\infty} & =&
 \frac {b_{1}^{3}} {b_{3}^{3}},
\label{eq:g1g2} 
\end{eqnarray} 
and
\begin{eqnarray}
T_{\infty} & = & \frac {4 \ b_{3}^{3}} {b_{1}^{3} + 2 \ b_{3}^{3}} = 
\frac {1-p_{\infty}^{3}} {1-q_{\infty}^{3}},
\label{eq:tinf} 
\end{eqnarray}
which clearly indicate that they are
similar to $(G_{2}/H_{2})_{\infty}$ and triaxiality parameter at
large $R$ of CT00, respectively.
Thus, the position angle $\Theta_{*}$ at large $R$ would also 
be identical to that of CT00 and can be calculated using 
equation (\ref{eq:tinf}) in (\ref{eq:pos}).

Likewise, at small $R$, the ratio $[(G_{2}+G_{2e})/(H_{2}+H_{2e})]_{o}$
 and the  triaxiality parameter $T_{o}$ are 
\begin{eqnarray}
\left(\frac {G_2+G_{2e}} {H_2+H_{2e}}\right)_{o} & =& 
\frac{\frac{12b_{1}^{3}}{b_{2}^{4}}-\frac{33.13a_{1}}{a_{2}^{3/2}}}
{\frac{12b_{3}^{3}}{b_{4}^{4}}-\frac{33.13a_{3}}{a_{4}^{3/2}}},
\end{eqnarray} 
and 
\begin{eqnarray}
T_{o} & = & \frac{4} {\left[2+ \frac{\frac{12b_{1}^{3}}{b_{2}^{4}}
-\frac{33.13a_{1}}{a_{2}^{3/2}}}{\frac{12b_{3}^{3}}{b_{4}^{4}}-
\frac{33.13a_{3}}{a_{4}^{3/2}}}\right]}.
\end{eqnarray}
It is clear that $[(G_{2}+G_{2e})/(H_{2}+H_{2e})]_{o}$ 
and $T_{o}$ differ much from $(G_{2}/H_{2})_{o}$ and 
triaxiality parameter at small $R$ of CT00, respectively, which 
implies that the position angle $\Theta_{*}$ at small $R$ would also be 
much different than that of CT00.

We have used  equation (\ref{eq:ba}) to calculate 
the axial ratios at asymptotic radii. In order to write 
it in some particular form, we define 
a few more functions  
\begin{eqnarray}
h_{4} & = & (6 \ h_{3})^{2} + 9 \ ( h_{1}-h_{2})^{2}, \nonumber \\
h_{5} & = & \frac {9} {4} \ ( h_{1} + h_{2})^{2}, \nonumber \\
h_{6} & = & 9 \ ( h_{1}^{2} - h_{2}^{2}) 
\label{eq:hfun}
\end{eqnarray}
of viewing angles $(\theta^{'}, \phi^{'})$ and then further 
define
\begin{eqnarray}
Z & \equiv & \left[h_{4} +h_{5} \ \left(\frac {G_{2}+G_{2e}}
 {H_{2}+H_{2e}}\right)^{2} + h_{6} \ \left(\frac {G_{2}+G_{2e}}
 {H_{2}+H_{2e}}\right)\right]^{{1}/{2}},
\label{eq:zform}
\end{eqnarray}
and
\begin{eqnarray}
A &=& \frac{\frac{12b_{1}^{3}}{b_{2}^{4}}-
\frac{33.13a_{1}}{a_{2}^{3/2}}}{\frac{12b_{3}^{3}}{b_{4}^{4}}-
\frac{33.13a_{3}}{a_{4}^{3/2}}}, \nonumber\\
B &=& \left\{(4\cos^{2}{\theta^{'}}\sin^{2}{2\phi^{'}})^{2}+
(3/2\sin^{2}{2\phi^{'}})^{2}A+(1+\cos^{2}{\theta^{'}})
\sin^{2}{\theta^{'}}\cos^{2}{2\phi^{'}}\right\}^{1/2}, \nonumber\\
\label{eq:baeo}
C &=& -\frac{2}{b_{o}}+\frac{2b_{1}^{3}h_{4}}{b_{2}^{4}}+
\frac{72b_{3}^{3}h_{5}}{b_{4}^{4}}+\frac{2a_{1}A}{a_{2}^{3/2}}
+\frac{12b_{3}^{3}}{b_{4}^{4}}B^{1/2}.
\end{eqnarray}

At very large $R$,  the axial ratio is given by  
\begin{eqnarray}
\left(\frac{b}{a}\right)^{2}_{\infty} &=& \frac{b_{o}^{3}-2b_{3}^{3}Z_{\infty}}
{b_{o}^{3}+2b_{3}^{3}Z_{\infty}},
\label{eq:baeinf}
\end{eqnarray}
where $Z_{\infty}$ is the value of $Z$ at very large $R$ which can
be evaluated by substituting the value of 
$[(G_{2}+G_{2e})/(H_{2}+H_{2e})]_{\infty}$ from (\ref{eq:g1g2}) 
in (\ref{eq:zform}), i.e., 
\begin{eqnarray}
Z_{\infty} & \equiv & \left[h_{4} +h_{5} \ 
\left(\frac {b_{1}^{3}} {b_{3}^{3}}\right)^{2} 
+ h_{6} \ \left(\frac {b_{1}^{3}} {b_{3}^{3}}\right)\right]^{{1}/{2}}.
\end{eqnarray}
On the other hand, at very small $R$, it can be expressed as
\begin{eqnarray}
\left(\frac{b}{a}\right)^{2}_{o} &=& \frac{C-12B\frac{b_{3}^{3}}
{b_{4}^{4}}}{C+12B\frac{b_{3}^{3}}{b_{4}^{4}}}.
\end{eqnarray}

Thus, we find that the position angle $\Theta_{*}$ of major axis and 
the axial ratio $b/a$ are the same as those of CT00 at large radii,
whereas they are much different at small radii. The reason behind 
this fact is that light from all radii 
will contribute to the projected properties evaluated at small radii
and hence, the effects of additional radial functions are seen there.
\section{Results}
\subsection{Radial profiles of the projected properties}
The  projected surface density $\Sigma$ of our considered 
triaxial mass model (\ref{eq:rhoe}) can be calculated
analytically. Its approximate elliptical isodensity 
contours show the variations in the axial ratio $b/a$ and
the position angle $\Theta_{*}$ of the major axis as 
functions of $R$. These features can be clearly seen from 
Fig. \ref{fig:fig1} where we present the radial profiles of $\Sigma$, 
$b/a$ and $\Theta_{*}$ for particular choices of 
intrinsic axial ratios,  $p_{o}=p_{\infty} \equiv p$ and 
$q_{o}=q_{\infty} \equiv q$, the viewing angles 
$(\theta^{'}, \phi^{'})$ and the constant parameters 
$(a_{1}, a_{2}, a_{3}, a_{4})$. On the basis of these results, 
our model can be considered as an example of simple analytical 
model showing ellipticity variations and isophotal twists in its 
projection along the line-of-sight. Since the resultant density 
distributing given in (\ref{eq:rhoe}) can have different 
contributions of the additional radial functions $g_{e}$ and $h_{e}$ 
for the various choices of constant parameters 
$(a_{1}, a_{2}, a_{3}, a_{4})$, in Fig. \ref{fig:fig2} we plotted 
the profiles of $b/a$ as a function of $R$ by allowing 
the different values of these constant parameters in such 
a manner that the resultant $\rho$ must remain positive. 
In this way, the effects of additional radial functions on 
the profiles of $b/a$ can be understood. We found that 
varieties of $b/a$ profile can be produced by changing 
the constant parameters $(a_{1}, a_{2}, a_{3}, a_{4})$ 
whose trends are found to be different from each other. 
Besides, it has also been noticed that the inclusions of 
additional radial functions produce small scale variations in 
$b/a$ profiles, specially at inner $R$, as compare to 
the smooth $b/a$ profiles found in case of CT00. 
Moreover, the profiles of position angle 
$\Theta_{*}$ also show small scale variation at inner $R$ for
different choices of the constant parameters 
$(a_{1}, a_{2}, a_{3}, a_{4})$ and 
in Fig. \ref{fig:fig3}, we have shown it only for one case which has 
all the intrinsic parameters similar to those considered 
in bottom right panel of Fig. \ref{fig:fig2}. The isodensity contours 
of many real elliptical galaxies show such small scale 
variations in the profiles of $b/a$ and  $\Theta_{*}$ 
as functions of $R$ rather than the smooth monotonically 
increasing or decreasing profiles of $b/a$ and  $\Theta_{*}$
 as those found in CT00. This 
suggests that the radial profiles of  $b/a$ and  $\Theta_{*}$ 
of our present traixial models, as a function of viewing angles 
$(\theta^{'},\phi^{'})$, can be compared with observations.

\subsection{Determination of observed projected properties exhibiting 
correlations}
\begin{figure}
\centering
\includegraphics[scale = 1., trim = 60 0 0 0, clip]{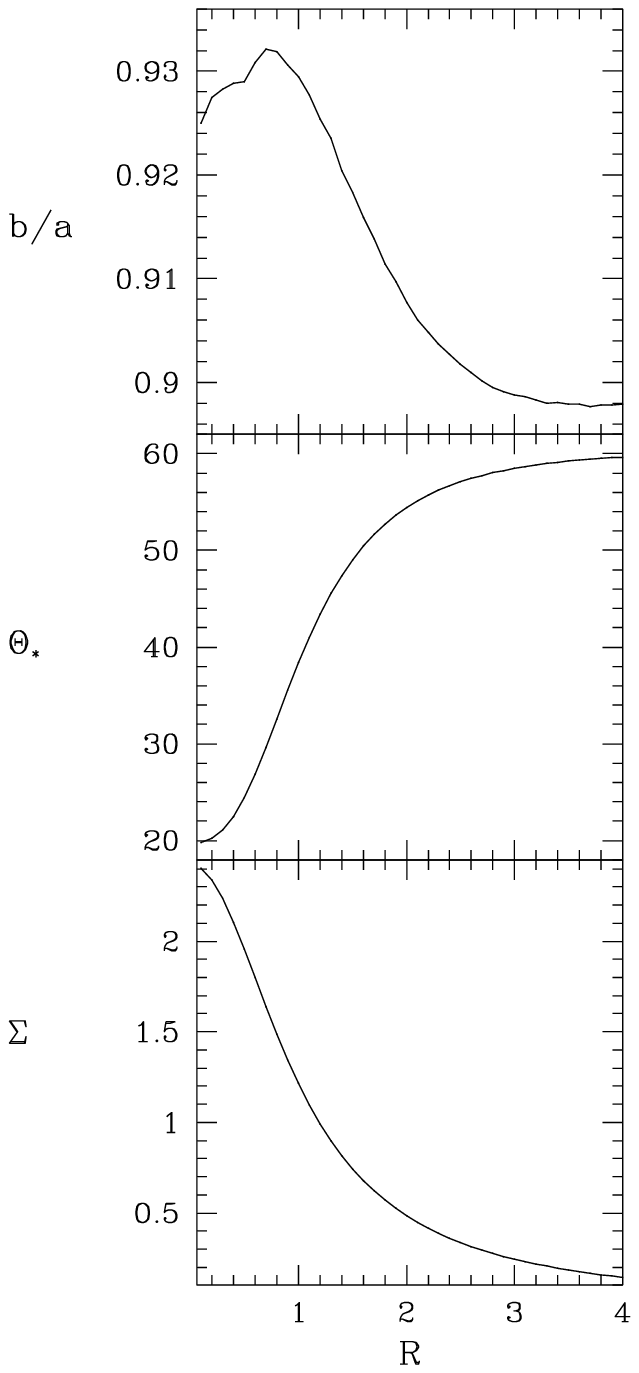}
\caption{\small Profiles of ${b}/{a}$, $\Theta_{*}$ and $\Sigma$ as 
functions of $R$ for a model with the intrinsic parameters 
$(p, q, \theta^{'}, \phi^{'})= (0.65, 0.60, 60^{o}, 10^{o})$ and 
the constant parameters $(a_{1}, a_{2}, a_{3}, a_{4})=
(0.2, 0.9, 0.2, 0.9)$. Here $\Theta_{*}$ 
is in degree, whereas $R$ and $\Sigma$ are 
in the units $b_{o}$ and $M/4 \pi b_{o}^{2}$, respectively.
Like $R$, the constant parameters $(a_{1}, a_{2}, a_{3}, a_{4})$
are also given in the units of $b_{o}$.}
\label{fig:fig1}
\end{figure}

\begin{figure}
\centering
\includegraphics[scale = 1., trim = 50 120 50 100, clip]{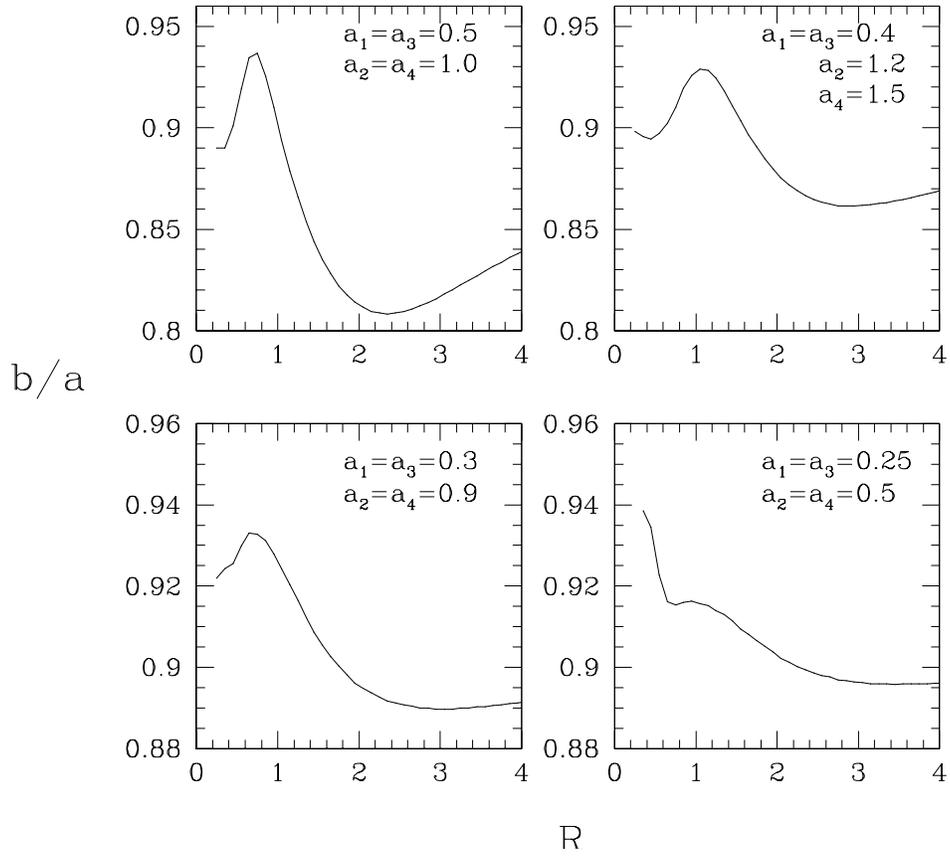}
\caption{\small Profiles of $b/a$ as functions of $R$ for four
 different choices of the constant parameters 
$(a_{1}, a_{2}, a_{3}, a_{4})$. The adopted values of  
$(a_{1}, a_{2}, a_{3}, a_{4})$ are given in top right corner
of each panel, whereas the remaining intrinsic parameters
of model, i.e., $(p, q, \theta^{'}, \phi^{'})$, are same as 
those in Fig $1$. Here $R$ and constant parameters 
$(a_{1}, a_{2}, a_{3}, a_{4})$ are given in the units 
of $b_{o}$.}
\label{fig:fig2}
\end{figure}

\begin{figure}
\centering
\includegraphics[scale = 1., trim = 80 120 50 100, clip]{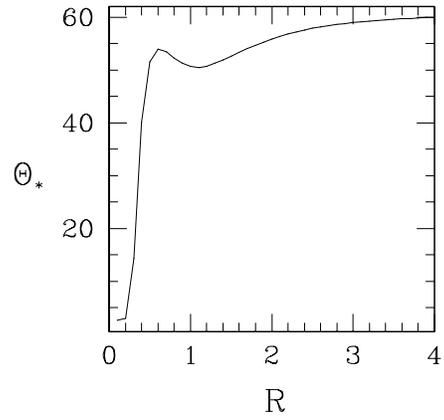}
\caption{\small Profile of position angle $\Theta_{*}$ as a function
of $R$. The intrinsic and constant parameters of model are 
same as those adopted in the bottom right panel of Fig $2$. 
Units of $R$, $\Theta_{*}$ and constant parameters 
$(a_{1}, a_{2}, a_{3}, a_{4})$ are same as those given in Fig. $1$.}
\label{fig:fig3}
\end{figure}

Determination of observed  parameters  which exhibit
non-overlapping and well-separated correlations for different 
choices of intrinsic axial ratios is desirable problem since these 
parameters  can be useful to get an estimate of 
intrinsic shapes of mass models, independent of viewing angles 
(SF94; TC01). It is also important to note here that 
correlation patterns between observed
projected properties evaluated at asymptotic radii, if exist, 
can save computing time due to availability of  
analytical expressions there. But, it would be meaningless to 
look for the correlations at asymptotic radii, since the observed
properties  at  asymptotic  radii  are  unobservable. With this
point in mind, we computed observed projected properties 
at $0.25$ and $4$ effective radii, $R_{e}$, in order to look for 
the above mentioned correlations at observable limits for the
five different sets of $(a_{1}, a_{2}, a_{3}, a_{4})$ which 
include one as in Fig. \ref{fig:fig1} and remaining four as in
 Fig. \ref{fig:fig2}. 
To compute $R_{e}$, we fitted the azimuthally average
surface density of our models to
$R^{1/4}$-profile \citep[][]{dev} between
the interval $0.25R_{e}\le  R \le4R_{e}$  and  
evaluated $R_{e}/b_{o}$ that depends on the intrinsic axial
ratios and viewing angles for each of five adopted sets of 
constant parameters $(a_{1}, a_{2}, a_{3}, a_{4})$. 
In the spherical limit of our model,
$R_{e}/b_{o}$ is found to be $\sim 4.4$. However, it is also
found that $R_{e}/b_{o}$ of our traixial models do not vary much
from its value calculated in the spherical limit. Even then, 
we computed $R_{e}/b_{o}$ for entire permitted range of 
intrinsic axial ratios and viewing angles 
for each set of $(a_{1}, a_{2}, a_{3}, a_{4})$ so that its 
exact value could be incorporated in calculations of correlations. 
Furthermore, we defined new parameter 
$(b/a)_{diff}=(b/a)_{.25R_{e}}-(b/a)_{4R_{e}}$, which exhibits 
ellipticity variation, and then found that 
the parameters $(b/a)_{4R_{e}}$ and  $(b/a)_{diff}$ are 
the better choice to represent the desired correlations. 

\begin{figure*}
\epsfxsize=6.5in
\epsfysize=6.5in
\epsfbox{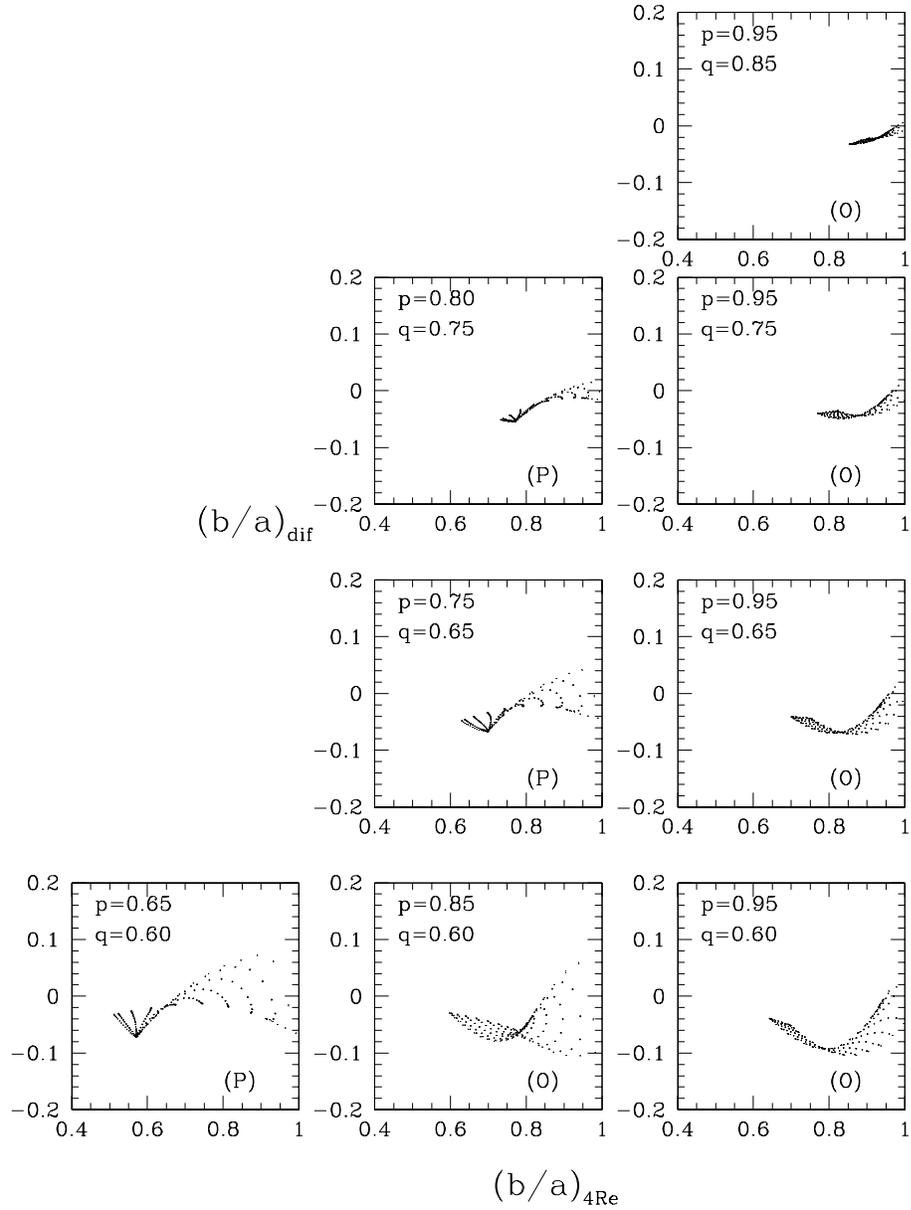}
\caption{\small Each panel shows correlation between 
$({b}/{a})_{4R_{e}}$ and $({b}/{a})_{diff}$, when a model
with a particular value of $(p, q)$ is
 projected at various viewing angles. Considered values of $(p, q)$
are indicated in top left corner of each panel, whereas constant 
parameters $(a_{1}, a_{2}, a_{3}, a_{4})$ are same as those adopted 
in Fig. $1$.}
\label{fig:fig4}
\end{figure*}

In Fig. \ref{fig:fig4}, we have shown the correlations between  
$(b/a)_{4R_{e}}$ and  $(b/a)_{diff}$ 
 for different choice of intrinsic axial ratios 
$p=p_{o}=p_{\infty}$ and $q=q_{o}=q_{\infty}$ roughly covering the
parameter space of $1\ge p \ge q \ge 0.5$ when our model with 
constant parameters $(a_{1}, a_{2}, a_{3}, a_{4})$ as in
 Fig. \ref{fig:fig1} is projected with varying viewing angles.
 We found that our model exhibits strong correlations which occupy
 well-separated regions in the parameters space of $(b/a)_{4R_{e}}$
 and  $(b/a)_{diff}$ for different choices of $(p, q)$. 

\begin{figure*}
\epsfxsize=6.5in
\epsfysize=6.5in
\epsfbox{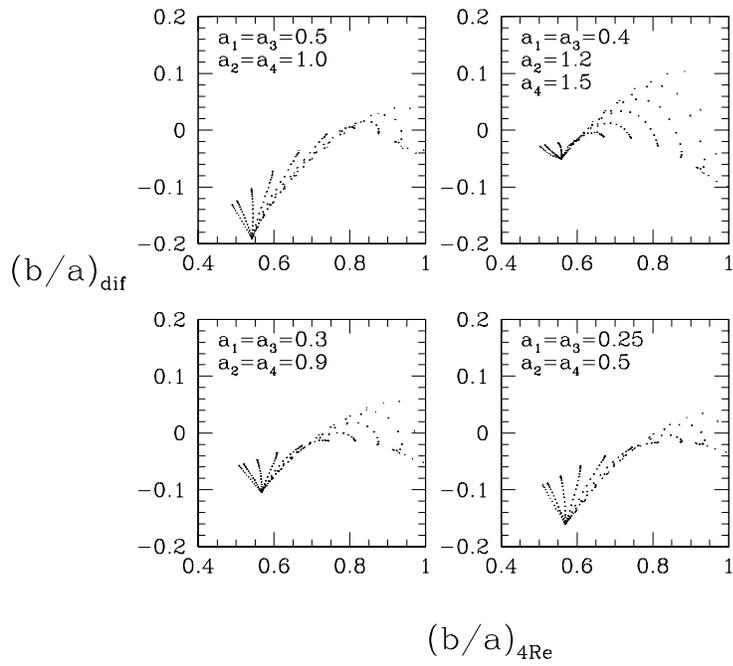}
\caption{\small Correlation between $({b}/{a})_{4R_{e}}$ and
 $({b}/{a})_{diff}$ for the fixed values of $(p, q)=(0.65, 0.60)$ but 
for different choices of constant parameters $(a_{1}, a_{2}, a_{3}, a_{4})$
 as indicated in top left corner of each panel.}
\label{fig:fig5}
\end{figure*}

\begin{figure*}
\epsfxsize=6.5in
\epsfysize=6.5in
\epsfbox{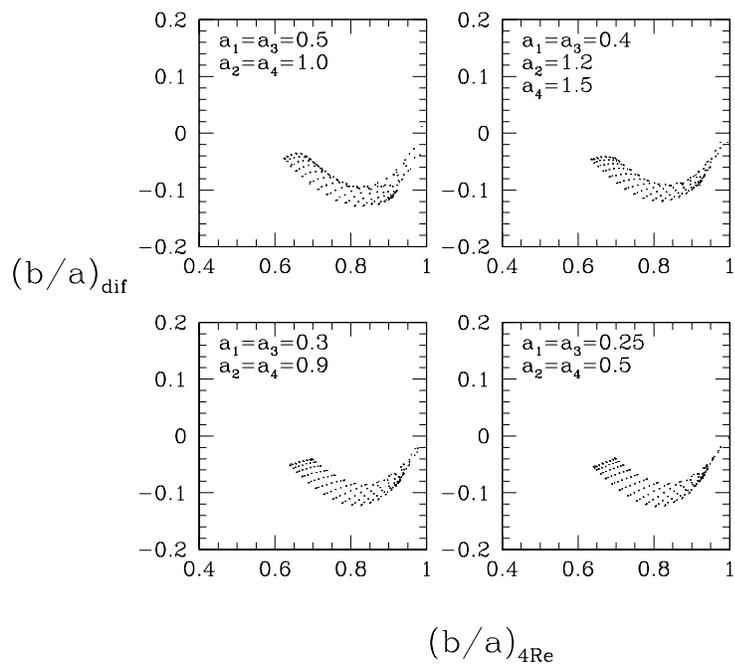}
\caption{\small Correlation plots as in Fig. 5 but for $(p,q)=(0.95,0.60)$.}
\label{fig:fig6}
\end{figure*}

It is interesting to note here that our model with remaining 
four values of $(a_{1}, a_{2}, a_{3}, a_{4})$ also show 
the qualitatively similar correlations as those found in 
Fig. \ref{fig:fig4}.
However, instead of considering all choices of  $(p, q)$ 
as in Fig. \ref{fig:fig4}, we have presented them for 
only two choices of $(p, q) = (0.65, 0.60)$ and 
$(p, q) = (0.95, 0.60)$ in Figs. \ref{fig:fig5} and \ref{fig:fig6},
 respectively. The former choice of $(p, q)$ represents highly 
prolate triaxial model, whereas the later choice of $(p, q)$ 
corresponds to highly oblate triaxial one.
Thus, these two choices of  $(p, q)$ are used to 
present non-overlapping correlations between 
a highly prolate and a highly oblate triaxial models that
can be seen by comparing each plot of Fig. \ref{fig:fig5} with that
 of Fig. \ref{fig:fig6}. 

Besides above found correlations where 
$\triangle p = |p_{o}-p_{\infty}|=0$ and 
$\triangle q = |q_{o}-q_{\infty}|=0$ are taken, it has also been seen 
that the correlation patterns still maintain as long as $\triangle p$ 
and $\triangle q$ are considered as 0.05. 

\begin{figure*}
\includegraphics[scale = 1., trim = 50 0 0 0, clip]{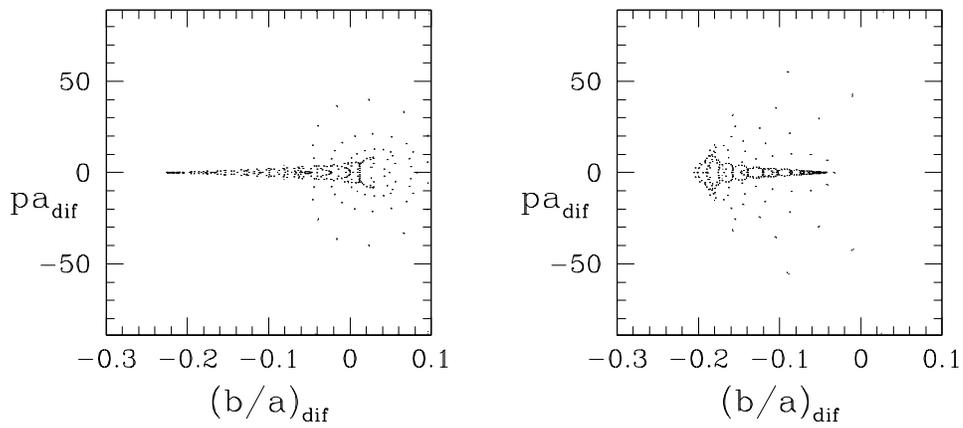}
\caption{\small Correlation plots between $({b}/{a})_{dif}$
 and $pa_{dif}$. The left panel represents correlation for model 
with $(p, q)=(0.65, 0.60)$, whereas the right panel shows it for 
$(p, q)=(0.95, 0.60)$. Constant parameters  $(a_{1}, a_{2}, a_{3}, a_{4})$
 are same as those adopted in Fig. $1$.}
\label{fig:fig7}
\end{figure*}

Finally, we have also examined the correlation plots by 
taking position angle
difference $pa_{dif} = (\Theta_{*o}-\Theta_{*\infty})$ as one 
of the observable parameters. Correlation plots between 
position angle difference $pa_{dif}$ and $(b/a)_{dif}$ for 
a highly prolate triaxial and a highly oblate triaxial are 
presented in Fig. \ref{fig:fig7} where constant parameters 
$(a_{1}, a_{2}, a_{3}, a_{4})$ are same as in Fig. \ref{fig:fig1}. 
Although some differences in the patterns between 
a highly prolate and a highly oblate triaxial are exhibited,
 they occupy overlapping regions. This suggests that 
$pa_{dif}$ is not a suitable observed parameter 
to develop well-separated correlations for two different 
choices of $(p, q)$ required for the shape determination.

But the non-overlapping correlations between 
the observed parameters $(b/a)_{4R_{e}}$ and  $(b/a)_{diff}$ 
are appeared to be useful in the prediction of the intrinsic
shapes of the mass models by mean of the graphical method of 
shape estimate of SF94. 

Despite the radial profiles of projected properties of our 
models are much different from those of CT00, 
its correlations between $(b/a)_{4R_{e}}$ and  $(b/a)_{diff}$ 
for different choices of $(p, q)$ are found to be 
qualitatively similar to those shown in Fig. $1$ 
of TC01 for CT00 model. This gives clear indication that
our models can be good candidate for generating 
the ensembles of models, by considering different choices 
of $\triangle p$ and $\triangle q$ for each set of constant 
parameters $(a_{1}, a_{2}, a_{3}, a_{4})$, which could be 
useful in the technique of TC01 in order to get the model 
independent shape estimate of a model galaxy. 
To get more insight over this issue,  an experiment to 
recover the intrinsic shape of a model galaxy is presented 
in the next section where we have used Bayesian statistics 
and exploited the correlations by employing 
the above mentioned observed parameters and ensembles of models.

\subsection{Recovery of the intrinsic shape of a model galaxy}

In order to get an estimate of the intrinsic shape of 
a model galaxy, we followed 
\citet[][hereafter, ST94a,b]{st94a,st94b} who used 
Bayesian statistics and considered photometric 
and kinematical data for estimating intrinsic shape of 
an individual galaxy. Moreover, he considered models of the
type $\rho = \rho(m^{2})$, which give rise to concentric and 
coaligned ellipsoids, and used a constant value of ellipticity 
for a galaxy under consideration.  

In our present study of the shape estimate of a model galaxy, 
we do not use any kinematical data. Rather, we have incorporated 
the variations in ellipticity by considering
$(b/a)_{4R_{e}}$ and  $(b/a)_{diff}$ as the observed parameters
which are calculated from our present model  by using 
following intrinsic parameters: $p_{\infty}$, $q_{\infty}$, 
$\triangle p$, $\triangle q$, $(\theta^{'}, \phi^{'})$ 
and $(a_{1}, a_{2}, a_{3}, a_{4})$. 
Here $p_{\infty}$ and $q_{\infty}$ are regarded as the shape 
parameters, while except of $(\theta^{'}, \phi^{'})$, the 
remaining intrinsic parameters are useful to generate ensembles of models. 

We launched a test case and considered an oblate traixial galaxy 
from our models with $p_{\infty}=0.95$, $q_{\infty}=0.60$, 
$\triangle p = 0$, $\triangle q =0$, $(\theta^{'}, \phi{'}) = 
(35^{\circ}, 10^{\circ})$ and constant parameters 
$(a_{1}, a_{2}, a_{3}, a_{4}) = (0.2, 0.9, 0.2, 0.9)$.
The observed parameters of this selected model galaxy are found to
be $(b/a)_{4R_{e}}=0.9398$ and  $(b/a)_{diff}=-0.0597$. 
We allowed a typical error of $0.01$ \citep[e.g.,][]{car} in
 ellipticities evaluated at $0.25 R_{e}$ and at $4 R_{e}$ which
 in turns set errors of $0.01$ and $0.014$ for $(b/a)_{4R_{e}}$ 
and $(b/a)_{diff}$, respectively. 

\begin{figure*}
\includegraphics[scale = 1., trim = 50 200 0 0, clip]{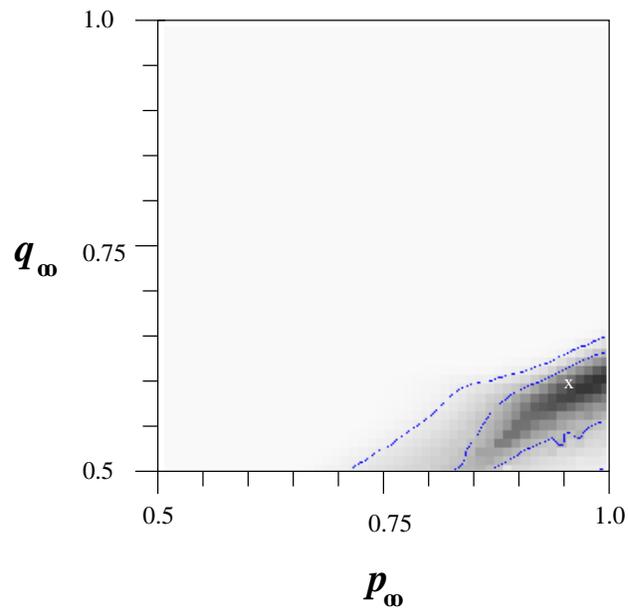}
\caption{\small Shape estimate of a model galaxy using ensembles of 
models. A white cross represents the true shape of a model galaxy. 
Inner and outer contours represent $1\sigma$ 
and $2\sigma$ error-bar regions, respectively.}
\label{fig:fig8}
\end{figure*}

In order to estimate the shape of this galaxy, we considered
our present models with five different sets of constant
parameters $(a_{1}, a_{2}, a_{3}, a_{4})$ as adopted in the
preceding sections. For each choice of 
$(a_{1}, a_{2}, a_{3}, a_{4})$, we choose 
$\triangle p \ge 0$ and $\triangle q \ge 0$ in such a way 
that for each $(p_{\infty}, q_{\infty})$, $(p_{o}, q_o)$ are
taken as $(\alpha p_{\infty}, \beta q_{\infty})$ where choices of
$(\alpha,  \beta)$ are $(1, 1)$, $(1.05, 1)$, $(0.95, 1)$,
$(1, 1.05)$, $(1, 0.95)$. This gives the ensembles of models for
a chosen value of $(p_{\infty}, q_{\infty})$. For each model 
with a particular value of $(p_{\infty}, q_{\infty})$ and above
given observed parameters and errors, we estimated
the likelihood $L$ of obtaining the observed data from the model and 
the posterior density $P$, which is the product of likelihood $L$ 
and the prior density $F$, by the formulas given in ST94a,b.
We assumed maximum ignorance of $F=(\sin \theta^{'}/4\pi) F_{s}$ 
by considering $F_{s}$=constant (cf. ST94b). We then integrated 
the posterior density $P$ over the viewing angles 
$(\theta^{'}, \phi^{'})$ and added it for different choices of 
$(\alpha,  \beta)$ and constant parameters 
$(a_{1}, a_{2}, a_{3}, a_{4})$. We plotted 
the resultant posterior density as a function of 
$(p_{\infty}, q_{\infty})$ which is shown by dark-grey shade 
in Fig. \ref{fig:fig8}. Here the darker shade corresponds to 
the higher posterior density. The inner and outer constant 
$P$ contours, enclosing
$68 \%$ and $95 \%$ of the resultant posterior density, 
show $1 \sigma$ and 
$2 \sigma$ error-bar regions, respectively. It is clear that
the true shape of a model galaxy, indicated by a white cross, 
lies within the $1 \sigma$ region. We also
found that the resultant posterior density is sharply peaked 
function of the shape parameters $(p_{\infty}, q_{\infty})$
since the $1 \sigma$ region is sufficiently narrow. This
satisfies the necessary condition of Bayesian statistics that 
the resultant posterior density to be relatively insensitive to the 
prior density in order to get the `likelihood dominated'
posterior density.
\section{Summary and discussion}

We have constructed  a family of triaxial models by inclusions of additional
 radial functions, each multiplied by a low-order spherical harmonic, in 
the models of CT00. This procedure of building triaxial models by
 addition of the additional radial functions was initially suggested by
 TC01 to enlarge the ensembles of models in their technique of
 intrinsic shapes determination of model galaxies. 
Projected surface density $\Sigma$ of our present triaxial models 
can be calculated analytically which allow us to derive the analytical
expressions of $b/a$ and $\Theta_{*}$ at asymptotic radii. However, 
the approximate elliptical isodensity contours of $\Sigma$ show 
the variations in $b/a$ and $\Theta_{*}$ as functions of $R$. 
This suggests that  our triaxial models can be considered 
as simple analytical models exhibiting ellipticity variations 
and isophotal twists. We have also studied the effects of 
the additional radial functions on the profiles of $b/a$ and $\Theta_{*}$ 
by considering different choices of constant parameters $(a_{1}, a_{2}, a_{3}, a_{4})$. 
In this regard, different trends of the profiles of $b/a$ and $\Theta_{*}$ 
are found for different contributions of 
additional radial functions. Furthermore, the profiles of 
$b/a$ and $\Theta_{*}$ show small scale variations at inner $R$ 
which are also seen in many real elliptical galaxies. Thus, our models 
can be useful to compare with observed photometric properties 
of real elliptical galaxies. 

TC01 pointed out that more 
ensembles of models, showing correlations between the observed parameters
for different choice of 
intrinsic parameters, are necessary to be explored to make 
their method of shape estimation more reliable. In this regard, our
models seem to be good candidate for providing another set of 
ensembles of models in the method 
of TC01, since result of shape estimation of a model galaxy using 
ensembles of our models is found to be satisfactory by utilizing 
the strong correlations between $(b/a)_{4R_{e}}$ and $(b/a)_{diff}$.
But it would be wise to look for larger sets of ensembles 
of models representing correlations between 
observed properties that could be useful in satisfactory 
determination of the shape of model galaxies before going 
ahead to apply this method of shape estimation to real ellipticals. 
\section*{Acknowledgements}
MD expresses her sincere 
thanks to PNU (Pusan National University, Busan, Korea) for 
providing financial support through the postdoctoral fellowship. 
PT would like to express his sincere thanks to 
ARCSEC (Astrophysical Research Center for the Structure and 
Evolution of the Cosmos, Seoul, Korea) for providing financial 
support through the postdoctoral fellowship which made 
this study possible. HBA thanks the Korea Science Engineering 
Foundation (KOSEF) for the support provided through 
grant No. R01-1999-00023.

\newpage

\end{document}